\RequirePackage[hyphens]{url}

\documentclass[conference,compsoc]{IEEEtran}

\usepackage{hyperref} 
\usepackage{tikz}
\usepackage{amsmath}
\usepackage{xspace}
\usepackage{paralist}
\usepackage{makecell}

\newcommand{\dbr}{{{\tt DBR}\xspace}}
\newcommand{\dbrs}{{{\tt DBR-s}\xspace}}

\usepackage{filecontents}
\newcommand{\etal}{\textit{et al.}}

\begin{document}

\date{}

\title{\Large \bf Let My Data Go: Data Brokers' Compliance 
with \\
CCPA-Mandated Opt-Out \& Deletion Requests}

\author{
Elina van Kempen, Gene Tsudik, Pragya Jhunjhunwala, and Mihir Raja\\UC Irvine
}

\maketitle

\begin{abstract}
Data brokers are (unfortunately) a largely American phenomenon dating back to 1980-s.  
They collect vast amounts of personal information about most adult U.S. consumers, 
mainly without their knowledge or consent. Accumulated data can be sold to anyone, including 
employers, landlords, insurance agencies, banks, governments  (local, state, federal, 
and even foreign), as well as various malicious actors. This, in turn, enables discrimination, 
surveillance, arrests, identity theft, and stalking. 

Recent regulations -- such as the California Consumer Privacy Act (CCPA) modeled 
after EU’s General Data Protection Regulation (GDPR) -- were introduced to bolster 
consumer privacy, e.g., the rights to: (1) opt-out of the sharing or selling one’s personal 
information, (2) delete one’s personal information, and (3) obtain a copy of that information. 
However, exercising these rights is not easy, as shown by our comprehensive study of 
the data broker ecosystem. We submitted both opt-out and deletion requests 
(using synthetic consumer identities, a choice that we believe is well-justified in this 
setting) under the CCPA to all California-registered data brokers and investigated their 
responses and lack thereof. 

While the majority seem to be compliant, a significant fraction is not and many failed 
to reply to (and/or acknowledge) consumer requests. Furthermore, some data brokers 
require intrusive consumer identity verification in order to exercise one’s opt-out rights, 
which is explicitly disallowed by the CCPA. There is also great disparity in the 
request submission process among data brokers as well as an extremely heavy 
(time and effort) overall consumer burden. This motivates an urgent need for 
streamlining and standardization of the consumer interface, stronger enforcement, 
and meaningful consequences for (especially sustained) non-compliance.
\end{abstract}


\section{Introduction \label{intro}}
Data brokers are commercial entities that collect a wide range of consumers' personal 
information, usually without the latter's knowledge or consent, and sell that
information to anyone willing to pay for it. It is estimated that there are over one 
thousand data brokers in the world, the majority of which are concentrated in the 
United States \cite{DataBrokersWatch, PRC}. Unfortunately, they represent a uniquely 
American phenomenon: the entire concept of data brokers
was ``born'' in the United States during the mid-to-late 1980-s. (Perhaps 
unsurprisingly, it originated in the state of Florida \cite{Funk}.)

Anyone can buy consumer information from data brokers, including private individuals, as 
well as foreign and domestic businesses and governments, including 
law enforcement and intelligence agencies. Data collected by data brokers 
range from relatively benign contact information to much more sensitive data, 
such as health, financial, and even precise geo-location as well as driving information
and shopping history, and lifestyle habits. 

In recent times, a more dangerous trend has emerged: the U.S. Department of
Homeland Security (DHS), particularly its Immigration and Customs 
Enforcement (ICE) branch, has been purchasing information about American
consumers from data brokers. This has been widely reported, e.g., by the
American Civil Liberties Union (ACLU) and the Electronic Frontiers Foundation 
(EFF) \cite{Tewari_Walter-Johnson}.
Although government agencies are not themselves allowed to collect data
about U.S. population, they are apparently not prohibited from buying it
from commercial entities.

If the above is not scary enough, in a 2025 killing spree in Minnesota, the 
perpetrator (who killed two people, including a state senator and her husband, 
plus wounded several others) collected extensive information on his intended victims
from about a dozen data brokers. Refer to the Wired article for details \footnote{
Minnesota Shooting Suspect Allegedly Used Data Broker Sites to Find Targets’ Addresses. \\
\url{https://www.wired.com/story/minnesota-lawmaker-shootings-people-search-data-brokers/}
}

Thus, data brokers pose a serious threat to consumers whose personal information is 
accumulated, sold, bought, and accessed by numerous entities, some of which certainly
do not have consumers' best interest in mind, to put it mildly.
Meanwhile, many (perhaps most) consumers are unaware of (1) who collects their 
personal information, (2) what exactly is being collected, and
(3) what, and to whom, their information is sold. This underscores the need for 
consumers to be given rights to control (e.g, at least via deletion and opt-out 
requests) collections and distribution of their sensitive/private information.

Recent privacy regulations around the world give consumers more agency over their 
personal information and grant them certain rights, e.g. access one's personal information,
delete it, or opt-out of its sale and sharing. Notably, these regulations  include
European Union's General Data Protection Regulation (GDPR) \cite{GDPR}, 
California Consumer Privacy Act (CCPA) \cite{CCPA}, and Brazilian Lei Geral de Proteção 
de Dados (LGPD) \cite{LGPD}, 

Furthermore, some regulations 
incorporate specific provisions for data brokers. In particular, California's 
Data Brokers Registration law \cite{DataBrokerLaw} mandates that every data broker 
doing business in the state much register annually and pay a registration fee. 
The resulting registry of registered data brokers is made publicly available \cite{brokerlist}.

Given such a public registry, consumers who reside in relevant jurisdictions can contact 
listed data brokers and exercise their legal rights. However, exercising one's rights can be 
complicated: contacting every single data broker is very time- and effort-consuming, since there are
many of them, plus methods of contact and user interfacees for submitting consumer requests 
vary widely. Moreover, getting a response to a consumer request is not guaranteed 
\cite{van2025consumer,DBLP:journals/popets/TakeYBGFMG24}. 

Very recent previous work analyzed compliance and behavior of data brokers when 
submitting consumer data access requests \cite{van2025consumer}, and found that more 
than 40\% did not respond to well-formed requests. Also, very few had data about the 
subject consumer, i.e., the researcher who conducted the study. Although an
access request yields important information about what kind of consumer information is 
collected by data brokers, delete and opt-out requests are (by far) much more numerous.
In 2024, there were 58 and 99 million, delete and opt-out requests, respectively, 
compared to just 4 million access requests. (These numbers are based on the 
aggregation of most recent reported metrics in the broker registry \cite{brokerlist}.) 
Therefore, compliance with delete and opt-out requests seems to be more 
important since they are much more frequent.

Motivated by the above, this paper reports on a comprehensive study of California-registered
data brokers in the context of delete and opt-out requests. Results of this study
can be viewed as a natural follow-on to \cite{van2025consumer}.
It aims to answer the following research questions: 
\begin{itemize}
    \item RQ1: What are the differences, if any, in the submission process 
    of opt-out and delete requests? For example, in terms of submission methods, 
    information requested, user interfaces, burden, etc.
    \item RQ2: To what extent do data brokers comply with CCPA in terms of delete
    and opt-out requests? Are there any trends in terms of differences in compliance with 
    one vs. the other?
    \item RQ3: What is the overall consumer burden involved in submitting each request type
    to all data brokers?
\end{itemize}
We note that delete and opt-out request types are semantically different: the latter 
requires no identity verification and is thus submittable by anyone, on anyone’s behalf.  
In contrast, deletion requests require identity verification. In the study described below,
We systematically probe data brokers with both request types using synthetic (non-existent) 
consumer identities and analyze their responses. Results show (1) significant 
non-compliance, (2) wide variety of requirements for submitting requests, 
(3) erroneous responses, and (4) very heavy burden for an average consumer to 
interact with data brokers.\\

\noindent This work aims to make three contributions:
\begin{enumerate}
\item Demonstrates a lack of transparency in data broker request processing, 
where the consumer is left unsure whether their request was even considered.
\item Yields evidence of recurrent CCPA non-compliance, whether due to
ignorance, negligence, or malfeasance. (We make no conclusion about specific 
non-compliance reasons).
\item Highlights the inordinate amount of effort a consumer must invest into 
dealing with numerous data brokers, via a range of submissions methods, and
a variety of user interfaces.
\end{enumerate}

\noindent {\bf Organization:}
The next section overviews relevant laws and current methods for 
the submission of consumer requests. Next, Section \ref{related}
summarizes prior work, followed by Section \ref{sec:methodology}
which describes the setup and preliminaries of the study. 
Then, Section \ref{results} discusses and analyzes study results. 
Next, Section \ref{limits} summarizes limitations and Section
\ref{future} outlines future work directions. Ethical considerations
are addressed in Section \ref{ethics}.

\section{Background}
This section describes the preliminaries, including relevant laws and regulations as well
as various means of submitting consumer requests. 

\subsection{Regulations}\label{sec:law}
Following the lead of European Union (EU) General Data Protection Regulation (GDPR) 
\cite{GDPR}, passed in 2016 and enforced since 2018, several privacy laws were passed 
in other jurisdictions. In particular, California Consumer Protection Act (CCPA) 
\cite{CCPA}, voted upon in 2018 and effective since 2020, affords California consumers 
rights over their personal information collected by business entities. CCPA itself 
is accompanied by after an opt-out requestCCPA Regulations \cite{CCR}. Consumer rights include:
\begin{itemize}
\item Deletion of personal information (CCPA \S 1798.105): This
means that the requesting consumer's data must be erased. Consumers can exercise this
right by submitting a ``Verifiable Consumer Request'' (VCR) (CCPA \S 1798.140(ak)). 
Businesses must verify the identity of the requesting consumer, 
usually by asking them to provide several items of personal information 
and matching these with collected data.
\item Opt-out of sale or sharing of personal information (CCPA \S 1798.120): 
This means that, although a business can retain consumer data in its databases,
it must stop all sharing and selling of that data. A consumer can exercise this right
by submitting a unique identifier which allows the business to find that consumer's data
in its databases. Businesses are explicitly not allowed to verify consumer 
identity for opt-out requests (CCPA Regs. \S 7026(c)), i.e., the opt-out
process is meant to avoid ``unnecessary friction.'' For example, in early 2026
the California Consumer Privacy Protection Agency fined Ford Motor Company
\$376K for unduly burdensome consumer opt-out process \cite{Ford}.
\end{itemize}
Personal information is defined as information that (CCPA \S 1798.140(v)): \\
\begin{quote} 
{\tt
``... identifies, relates to, describes, ..., or could reasonably be linked, directly 
or indirectly, with a particular consumer or household...''
}\\
\end{quote}
Once a business receives a delete or opt-out request, it must, respectively:
\begin{itemize}
\item Delete any personal information that relates to the consumer and 
respond to the consumer with the output of the request, within 45 calendar days (CCPA \S  1798.130(a)(2)), 
e.g. data deleted, or no data found, or impossible to verify identity.
\item Stop selling and sharing of the consumer's personal information within 
15 business days and wait 12 months before asking again for the consumer's 
consent to share their information (CCPA \S 1798.135). 
\end{itemize}
For opt-out requests, CCPA Regulations state that there must be an explicit confirmation of a 
request having been processed (CCPA Regs. \S 7025(b)): \\
\begin{quote} 
{\tt
``...a business must provide a means by which the consumer can confirm that their request 
to opt-out of sale/sharing has been processed by the business. For example, the business 
may display on its website “Opt-Out Request Honored” in accordance with section 7025, 
subsection (b)(6), and display in the consumer's privacy settings through a toggle or 
radio button that the consumer has opted out of the sale/sharing of their personal information.''
}\\
\end{quote} 

For deletion requests, CCPA Regulations state that (CCPA Regs. \S 7023(b)): \\
\begin{quote}
{\tt
``In responding to a request to delete, a business shall inform the consumer whether 
it has complied with the consumer's request.''
}\\
\end{quote}
In addition to CCPA requirements, the Data Broker Registration law \cite{DataBrokerLaw} 
mandates yearly registration for all data brokers operating in the state of California. 
A data broker is defined as: \\
{\begin{quote} {\tt
``... a business that knowingly collects and sells to third parties the personal 
information of a consumer with whom the business does not have a direct 
relationship.''
}\\
\end{quote}
The list of all California registered data brokers is publicly available 
on the CCPA website \cite{brokerlist}.

Besides from California, a few other states mandate a data broker registration. 
For example, Vermont \cite{Vermont}, Oregon \cite{OREGON} 
and Texas \cite{Texas} as well as New Jersey and Connecticut have
introduced data broker registration bills and established registries.
However, none is as comprehensive or stringent as the California's CCPA.
Notably, there is also a recent U.S. Federal bill that
includes mandating data broker registration \cite{SECURE}.

\subsection{Submitting Delete \& Opt-Out Requests in Practice}
We now overview the current means of submitting consumer requests to data brokers.
{\bf NOTE:} from here on we use \dbr\  (\dbrs) as a short-hand for data broker(s).

\subsubsection{Direct Contact} 
Consumers can directly contact \dbrs\ to submit requests. Each \dbr, in its privacy policy, 
must specify at least 2 ways to contact them \cite{CCPA}. While this method is very tedious 
and time-consuming, it offers the most granularity and
flexibility in terms of how and which \dbrs\ are contacted.

\subsubsection{3rd-Party Services} 
Certain for-profit services offer to submit requests on behalf of the consumer for a fee. 
These entities include: DeleteMe \cite{deleteme}, Aura \cite{aura}, Incogni \cite{incogni}, and 
Optery \cite{Optery_2025}. 
While doing this is convenient for consumers, it can be expensive, and, according to recent
findings, \dbr\ coverage and data removal success rate are unimpressive \cite{HePets25}. 
Also, the method entails giving one's personal information to yet another 3rd party of potentially
dubious trustworthiness, i.e., it may get hacked or may itself
engage in collecting and/or selling consumer data.

\subsubsection{The CCPA DROP System (Portal)\label{drop}} 
On January 1 2026, California Privacy Protection Agency (CPPA) launched the 
{\it Deletion Request and 
Opt-out Platform (DROP)} site~\cite{DROP}. California-resident consumers can 
use DROP to make a unified
delete or opt-out request for {\bf all} registered \dbrs, by providing several items of 
identifying information. So-called ``basic'' information includes:
{\it Name(s), Date of Birth, ZIP code(s), Email address(es), and Phone number(s)}, while 
optional information includes:  {\it Mobile Advertising IDs (MAIDs), Connected TV IDs, and Vehicle 
Identification Numbers (VINs)}. Data brokers were also mandated to 
create DROP portal accounts by that date. Nonetheless, they were not required to begin
processing consumer requests until August 1, 2026.

However, even though the DROP portal is available in full from August 1, 2026,
its efficacy, including \dbrs' compliance, and usability
are unknown as of now -- mid-November 2026. This is because \dbrs\ are allowed up to 45 days 
to pull consumer requests and another 45 days -- to act upon them, i.e., delete/opt-out all personal 
information for any consumer whose identifier is stored at the portal \cite{DROP,deleteact2026}.

Furthermore, since the DROP portal is restricted to {\bf current} 
California residents, consumers in other jurisdictions (including former California residents)
must contact \dbrs\ directly or rely on third-party services.
Another issue is that DROP is aimed at {\bf living adults}. It is well-known that
\dbrs collect and retain information about deceased individuals. DROP explicitly states
that this case is not supported. Worse yet, \dbrs routinely collect
information about children/minors. In this case (as in the case of elderly and/or incapacitated
individuals), DROP requires the use of an "authorized agent" which is defined as a parent,
legal guardian, or a third-party commercial service.

\section{Related Work \label{related}}
Relevant prior work is reviewed in this section is summarized in Table \ref{tab:related}.

Many related studies \cite{DBLP:conf/soups/MartinoRWQLA19, DBLP:journals/popets/MartinoMQAL22, 
DBLP:conf/apf/BonifaceFBLS19,wong2019right,DBLP:conf/icws/BufalieriMMS20,DBLP:conf/IEEEares/KrogerLH20,
DBLP:conf/sicherheit/HerrmannL16} examined the behavior of popular websites and businesses by 
submitting GDPR requests, with response rates ranging between 55\% and 93\%, and underlined 
privacy issues of submitting data requests. 

Several studies created synthetic or pseudonymous persons and entered their information into websites 
or apps, thus ensuring that the latter collected information about the synthetic persons 
\cite{DBLP:journals/popets/SamarinKSBYWAFHE23,DBLP:conf/sicherheit/HerrmannL16,DBLP:conf/icws/BufalieriMMS20}. 
The same would be very complicated to do with \dbrs\ since they are not consumer-facing entities.
Other work specifically contacted companies for which data was previously collected about the 
data subjects, i.e., the researchers themselves \cite{DBLP:conf/soups/MartinoRWQLA19,DBLP:journals/popets/MartinoMQAL22,wong2019right}.

To evaluate the possibility of privacy-preserving requests, Adhatarao \etal \cite{adhatarao2021ip} 
submitted requests using only an IP address as proof of identity. All requests were subsequently denied, 
since companies decided that they could not verify an identity with only an IP address. 

In 2023, Samarin \etal \cite{DBLP:journals/popets/SamarinKSBYWAFHE23} submitted CCPA ``right-to-know'' 
(i.e., data access) requests to 109 Android app developers and reported their findings. The overall 
non-response rate was 19\%. The same study showed alarming under-reporting of many types of 
collected information.

Take \etal \cite{DBLP:journals/popets/TakeYBGFMG24} submitted GDPR and CCPA requests to 20 people 
search websites -- a category of \dbrs\ that make personal information easily accessible 
online through a searchable database. (Theses sites are often the ones that pop up first in the
list when doing a Google search on a person's name.) Results showed that submitting requests 
was not straightforward, and the process varied even across requests made to the same \dbr,
depending on who was submitting them.

Recently, He \etal \cite{HePets25} analyzed the efficacy of ten PII removal services (paid services 
that claim to remove their customers' personal information from \dbrs' sites), and found that coverage 
varied greatly, with none being fully effective at removing personal information from \dbrs.

Finally, Van Kempen \etal \cite{van2025consumer} investigated 
compliance with consumer data access requests. In that study, researchers submitted 
requests using their own personal information to all registered \dbrs\ and found that over 
40\% were unresponsive. In contrast, this paper focuses on much more popular 
deletion and opt-out requests. 

\begin{table*}[!h]
\centering
\caption{Related Work Summary.}
\label{tab:related}
\resizebox{\textwidth}{!}{%
\begin{tabular}{|ccccccc|}
\hline\cline{1-7} 
& & & & & & \\
  \textbf{Result/Study} & \textbf{Law} & \textbf{Request type} & \textbf{Resp rate} & \textbf{Consumer} & \textbf{Entities studied} & \textbf{Year(s)} \\
& & & & & & \\
  \hline \cline{1-7}
  {\bf This Work} & CCPA & Delete/Opt-out & 30-86\% & Synthetic & Data Brokers & 2025-2026 \\ \hline
  Van Kempen \etal \cite{van2025consumer} & CCPA & Access & 57\% & Real & Data Brokers & 2024-2025 \\ \hline
  Take \etal \cite{DBLP:journals/popets/TakeYBGFMG24} & CCPA/GDPR & Access/Delete & 82-100\% & Real & People Search Websites & 2024 \\ \hline
 Samarin \etal \cite{DBLP:journals/popets/SamarinKSBYWAFHE23} & CCPA & Access & 81\% & Synthetic & Smartphone apps & 2023 \\ \hline
 Adhatarao \etal \cite{adhatarao2021ip} & GDPR & Access &  57\% & Real (IP addresses) & Popular Websites & 2021 \\ \hline
 Martino \etal \cite{DBLP:journals/popets/MartinoMQAL22} & GDPR & Access & 93\%  & Real & Popular Websites & 2021 \\ \hline
Bufalieri \etal \cite{DBLP:conf/icws/BufalieriMMS20} & GDPR & Access & 71\%  & Pseudonymous/Real & Popular Websites & 2020 \\ \hline
Martino \etal \cite{DBLP:conf/soups/MartinoRWQLA19} & GDPR & Access & 93\% & Real & Popular Websites & 2019 \\
 \hline
  Kroger \etal \cite{DBLP:conf/IEEEares/KrogerLH20} & GDPR & Access & 77-83\% & Real & Smartphone apps & 2015-2019 \\ \hline
Urban et al \cite{DBLP:conf/esorics/UrbanTDHP19} & GDPR & Access & 58\% & Real & Tracking services & 2018 \\ \hline
 Wong \etal \cite{wong2019right} & GDPR & Portability & 75\% & Real & Popular Websites & 2018\\ \hline
 Herrman \etal \cite{DBLP:conf/sicherheit/HerrmannL16} & GDPR & Access & 55-77\% & Synthetic & \makecell{Smartphone Apps \\ \& Popular Websites} & 2016 \\
\hline
\end{tabular}%
}
\end{table*}

\section{Methodology \label{sec:methodology}}
This section described our study methodology, including timeline, identities, 
\dbrs\ selection/exclusion, and ethical considerations.

\subsection{Study Timeline}
We submitted opt-out and deletion requests to all \dbrs\ registered in the state of 
California, as of early October 2025. The bulk of the submission process lasted roughly two months, 
starting in early October, and ending in early December, 2025.
Two researchers, based in California, split the work: one was tasked with submitting delete, 
and the other -- opt-out, 
requests. Each researcher worked on their own schedule and at their own speed. Consequently, 
the rate of request submission varied greatly between the two researchers and their
individual request submission timelines were not in any way dependent on each other.
This was done for two reasons: (1) synchronizing their workloads would have been
infeasible due to external factors, and (2) independent timelines naturally inhibit 
correlation between their respective activities, as would be observed by \dbrs.

The responses started trickling in from the very beginning. 
Most \dbrs\ answered either on the day of, or a few days after, receiving the 
request. Latest answers were received about 60 days after delete, and about 100 days
after opt-out, requests were submitted. Further response timeline detailed are
presented in Sections \ref{delpro} and \ref{optpro}.

Note that the discussion of overall time it took to submit both request types
appears later in Section \ref{timetaken}.

\subsection{Use of Synthetic (Fictitious) Identities \label{ident}}
We used two distinct synthetic (non-existent or fictitious) identities, one for each request type. 
We believe that synthetic identities are preferable to real ones, since submitting requests (to 
hundreds of \dbrs) using the information about real people would impact those individuals. 

Also, unlike closely related recent work \cite{van2025consumer}}, we chose not to use the 
researchers' (i.e., this study and paper authors') own information and identities, 
since a simple web search on  
researchers' names would hint at their affiliations, publication histories, and research profiles. 
That would likely expose the purpose of this study, which would have been obviously undesirable:  
we wanted \dbrs\ to be unaware of the study's purpose, since otherwise they could treat
our requests differently (whether better or worse) than organic ones.

Another alternative could have been to recruit other individuals (i.e., non-authors), such as
local university students, and ask them to submit requests using their own information. In that case,
obtaining IRB approval would have been nearly impossible. (This is based on actual communication
with our IRB personnel.) Furthermore, it would naturally trigger serious ethical concerns, even 
with IRB approval.

We used two identities, as opposed to just one, since we wanted the delete and opt-out 
requests not to be easily correlated by \dbrs. Of course, we could have used more than two identities,
in the extreme best-case scenario -- one distinct identity per request. However, given 
the large number of queries, creating as many realistic-sounding identities along with 
accompanying data (dates-of-birth, addresses, phone \#-s, email) would have been prohibitively burdensome.

Exact details of the two synthetic identities are shown in Tables \ref{tab:synthI} and \ref{tab:synthII}.
The names correspond to non-existent individuals. We carefully selected the names by thoroughly
searching for each and made sure that these names do not correspond to any living (or dead) 
individuals, in order to avoid any impact on real persons. 

As reflected in the tables, dates-of-birth and
phone numbers are fictitious, though their area codes match to those assigned to the geographical
areas that correspond to the addresses we used.  However, the email addresses are real in the sense 
that we created them in order to be able to confirm email account access if that was needed during 
the request submission process. Finally, Mobile Advertising IDs (MAIDs) are fake. A MAID is a 
sequence of 32 hexadecimal digits generated by the OS of a mobile device (e.g., a smartphone or tablet)
used by advertisers to track activity, app usage, and location data, even across devices.
It is resettable by users.

We chose one female- and one male-seeming identity for the sake of gender diversity. We also opted 
for having a substantial age gap between them (over 30 years), for age diversity. Physical addresses 
included Northern (Sacramento) and Southern (Riverside) California, for geographical diversity. 
Finally, the types of physical addresses that we used were motivated by socioeconomic diversity: 
the Riverside address corresponds to a real trailer park, while the (real) street name in 
Sacramento is located in a solidly middle-class neighborhood. In both addresses,
zip-codes are real.

\begin{table}[h] \sf
\caption{Synthetic Identity I (for deletion requests).}
\centering
\label{tab:synthI}
\resizebox{\linewidth}{!}{%
\large \begin{tabular}{| l | c | }
\hline\cline{1-2}
{\bf\large Item } & {\bf\large Value }   \\ \hline\cline{1-2}
Name  & Patrick Puzzolente
\\ \hline
Email &  patrickpuzzolente@gmail.com
\\ \hline
Date-of-Birth  & July 7, 1993 
\\ \hline
Phone \# (fake) &  951-893-3150
\\ \hline
Address &  220 E. La Cadena Dr, \\
(real, trailer park) &  Riverside, CA 92507\\ \hline
MAID (fake)& F2E4C0A7-11B8-4009-A6D4-3C9B7F0E2A11
\\ \hline\cline{1-2}
\end{tabular}%
}
\end{table}

\begin{table}[h] \sf
\caption{Synthetic Identity II (for opt-out requests).}
\centering
\label{tab:synthII}
\resizebox{\linewidth}{!}{%
\large \begin{tabular}{| l | c | }
\hline\cline{1-2}
{\bf\large Item } & {\bf\large Value }  \\  \hline\cline{1-2}
Name  & Annamaria Krahenschrei
\\ \hline
Email &  annamariakrahenschrei@gmail.com
\\ \hline
Date-of-Birth  & May 24, 1960 
\\ \hline
Phone \# (fake) & 916-646-2451 
\\ \hline
Address & 6452 Valetta Way, \\
(fake \#, real street) & Sacramento, 95820\\ \hline
MAID (fake)& 6A3F1B9E-0D2C-4A51-9E3B-8F2C1A7D9B4E
\\ \hline\cline{1-2}
\end{tabular}%
}
\end{table}

\subsection{Delete and Opt-out Email Templates \label{app:emails}}
The opt-out and delete templates are shown in Figures 
\ref{fig:out-template} and \ref{fig:del-template}, respectively.

\begin{figure}[h]
    \centering \tt
    \caption{Opt-out request email template}
    \vspace{0.3cm}
    \label{fig:out-template}
    \resizebox{\linewidth}{!}{%
    \begin{tabular}{|l|}
    \hline 
    Hello,\\
    \\
    Under the CCPA opt-out rights, \\
    I hereby request that you stop \\
    selling or sharing my personal \\
    information. Please let me know if \\
    you need additional information. \\
    \\
    Thank you for processing this request.\\
    \\
    Sincerely,\\ 
    Annamaria Krahenschrei
    \\ \hline
    \end{tabular}%
    }
\end{figure}
\begin{figure}[h] 
    \centering \tt\large
    \caption{Delete request email template}
    \vspace{0.3cm}
    \label{fig:del-template}
    \resizebox{\linewidth}{!}{%
    \begin{tabular}{|l|}
    \hline \large
    \noindent
    To whom it may concern,\\
    \\
    I am writing to request the deletion \\
    of my personal information under the \\
    CCPA. Please let me know if you need \\
    any further information. \\
    \\
    Thank you for your timely response.\\
    \\
    Best regards, \\
    Patrick Puzzolente
    \\ \hline
    \end{tabular}%
    }
\end{figure}

\subsection{\dbr\ Selection/Exclusion \label{exclude}}
Using the California Data Broker Registry \cite{brokerlist}, we aimed to contact all 
$535$ \dbrs\ registered at the time of the study, in late 2025. However, some \dbrs\ had
to be excluded from the study due to inability to submit requests, for various reasons, 
such as:
\begin{itemize}
    \item Website issues, mainly due to broken URLs
    \item Duplicate registrations of the same business
    \item Requirements to provide information (for identity verification purposes) 
    that we either could not, or did not want to, invent, e.g., cookie values or Social Security N
    umbers (SSNs) 
    \item Issues with forms that prevented successful submission, e.g., fields or CAPTCHA-s 
    that were mis-configured.
    \item Lack of contact information in the privacy policy, either for all requests or 
    specifically missing methods for deletion requests.
    \item Other: this includes individual errors, e.g., being forced to log into an account, 
    or email bouncing back.
\end{itemize}

We wound up submitting a total of $322$ deletion, and $358$ opt-out, requests. These numbers 
correspond to successfully contacted \dbrs. The difference ($36=358-322$) is, for the most part, 
due to the fact that some \dbr\ websites only had opt-out forms available and omitted deletion rights. 
In addition, CCPA, as mentioned earlier, mandates no identity verification for opt-out requests, 
while it is is required for deletion requests: we did not submit to \dbr\ asking for PII we could not 
invent. Therefore, we could submit $36$ more of opt-out requests. Table \ref{tab:exclusions} 
summarizes \dbr\ exclusions.

\begin{table}[h]
\centering
\caption{\dbrs Excluded from the study.}
\label{tab:exclusions}
\large\begin{tabular}{|l | c | c|}
\hline\cline{1-3}
{\bf\large Exclusion Reason} & {\bf\large \#\dbrs } & {\bf\large \#\dbrs}  \\ 
{\bf\large } & {\bf\large (delete)}  &  {\bf\large (opt-out)} \\ 
\hline\cline{1-3}
Website down & 8 & 14 \\ \hline
Duplicate & 25 & 25 \\ \hline
Asked too much PII & 85 & 82 \\ \hline
Form issue(s) & 33 & 46 \\ \hline
No contact info given & 45 & 7 \\ \hline
Other & 17 & 3\\
\hline\cline{1-3}
\end{tabular}
\end{table}


For each \dbr, we accessed the link indicated in the \dbr\ registry that contains details on how consumers 
can exercise their CCPA rights, including how to delete their personal information. This link usually leads 
to one of: (1) a \dbr's privacy policy, or (2) a CCPA-specific privacy policy, or (3) a form where the 
consumer can  exercise their privacy rights. We submitted deletion and opt-out requests by form whenever 
possible. Whenever no form was available, we submitted requests by email. The email templates that we used are 
shown in Section \ref{app:emails}. The two email templates are differently worded, for obvious reasons.

We recorded dates of all request submissions, confirmation of request receipt by \dbrs\ (if any) and their 
final responses. These responses were later categorized and analyzed.

\subsection{Ethical Considerations \label{ethics}}
The authors' Institutional Review Board (IRB) deemed this study to fall into the 
``Non-Human Subject Research'' category.  

Each \dbr\ not excluded from the study
(see Section~\ref{exclude}) was contacted only once per request type (twice total).

One potential issue is our use of two fictitious identities (along with related 
information for each) in request submission. This can be viewed as a kind of 
deception. However, as mentioned in Section \ref{ident},
we made sure that these identities did not correspond to any real person, whether 
living or dead. Also, their names and other information were carefully constructed to
appear realistic. We believe that this was justified, since the alternatives (e.g.,
using identities of real people, including those of authors) would have been 
ethically problematic and/or would have leaked the purpose of submitted requests.
Also, we did not invent any sensitive personal information when it was required, e.g., 
SSNs or copies of government-issued ID cards. 
Instead, we opted to exclude \dbrs\ that requested such information. 

Finally, as part of the IRB application, our intended use of fictitious identities 
was fully disclosed.

\section{Results \& Analysis}\label{results}
We now present study results obtained from submitting requests to \dbrs\ 
as described in Section \ref{sec:methodology}.

\subsection{Results Summary: Compliance Rates} \label{sec:compliance}
We observed the following instances of non-compliance, summarized in Table \ref{tab:res-sum}:
\begin{enumerate}
\item Lack of response after a request to delete: as required by CCPA Regulations \cite{CCR} 
and mentioned in Section \ref{sec:law}, \dbrs\ must notify the consumer of the status of 
their request for deletion after processing it. We found that a majority of \dbrs\ did not 
answer after an email-based request submission, and a majority of \dbrs\ considered the  
confirmation page (after form submission) as a response.
\item Identity verification during opt-out requests: again, CCPA Regulations \cite{CCR} specify
that businesses should not verify the identity of consumers when they submit opt-out requests 
(see Section \ref{sec:law}). Recently, Ford Motor Company was fined \$375,703 due to excessive 
identity verification in their opt-out request submission process. Excessive verification 
involves asking for more information that is actually needed to identify the consumer in the 
\dbr's database \cite{Ford}. It occurs when \dbrs\ ask for too many items
of personal information, or attempt to verify consumer's access to their email address 
or phone number.
\item Late answers deletion requests: \dbrs\ must process the request and respond within 
45 calendar days. They can request an extension for another 45 days, however, 
none did so in this study.
\end{enumerate}
Furthermore, we observed the following instances of potential non-compliance summarized 
in Table \ref{tab:res-sum}:
\begin{enumerate}
\item Lack of answers after opt-out requests: while automatic form submissions can be 
considered a confirmation of opt-out, \dbrs\ contacted by email and who did not reply 
did not show any confirmation that the opt-out request was taken into account.
\item Late answers after opt-out requests: \dbrs\ must process any opt-out request within 
15 business days. Some \dbrs\ responded to our request after that deadline: it is unlikely 
that the personal information would be opted-out before the 15 business days deadline.
\item Potentially excessive identity verification: While these \dbrs\ did not specify 
that they needed to verify the consumer's identity, they asked for a greater amount of 
personal information that is necessary for identifying the consumer for opt-out requests.
\item Erroneous answers for deletion requests: \dbrs\ are supposed to announce actions 
taken to the consumer after processing a deletion request. Many \dbrs\ replied that 
data was deleted, even though the consumer was synthetic and the data obviously does not 
and did not exist. 
\item Requiring access to email address to learn the status of an opt-out request: while some 
\dbrs\ (correctly) did not require a consumer to verify access to their email address when 
submitting an opt-out request, \dbrs\ that use the OneTrust platform need the consumer to 
verify access to their email address before being able to view the response. While the personal 
information is already opted-out, this introduces an additional burden on the consumer if 
they want to check the \dbr's answer.
\end{enumerate}

\begin{table}[]
\label{tab:res-sum}
\begin{center}
\large\begin{tabular}{|l|l|l|} 
\hline
\multicolumn{1}{|l|}{\textbf{\large Identified Issue}}              & \multicolumn{1}{l|}{\textbf{\large Req Type}} & \textbf{\large \%-age} \\ \hline
\multicolumn{3}{|c|}{\textbf{Non-compliance}}                                                            \\ \hline
\multicolumn{1}{|l|}{Lack of answer}              & \multicolumn{1}{l|}{Deletion}         & 70\%                \\ \hline
\multicolumn{1}{|l|}{Identity verification}       & \multicolumn{1}{l|}{Opt-out}          & 22\%                \\ \hline
\multicolumn{1}{|l|}{Late answer}                 & \multicolumn{1}{l|}{Deletion}         & 1.2\%               \\ \hline 
\multicolumn{3}{|c|}{\textbf{Potential non-compliance}}                                                  \\ \hline
\multicolumn{1}{|l|}{Lack of answer}              & \multicolumn{1}{l|}{Opt-out}          & 14\%                \\ \hline
\multicolumn{1}{|l|}{Late answer}                 & \multicolumn{1}{l|}{Opt-out}          & 12\%                \\ \hline
\multicolumn{1}{|l|}{Identity verification}       & \multicolumn{1}{l|}{Opt-out}          & 39\%                \\ \hline
\multicolumn{1}{|l|}{Erroneous answer}            & \multicolumn{1}{l|}{Deletion}         & 2.5\%               \\ \hline
\multicolumn{1}{|l|}{Email verif. to view answer} & \multicolumn{1}{l|}{Opt-out}          & 11\%                \\ \hline
\end{tabular}
\end{center}
\caption{Instances of Non-compliance}
\end{table}

\subsection{Request Submission: Lack of Uniformity}\label{sec:nostandard}
Previous work \cite{van2025consumer} has shown that submitting access requests is very time-consuming and 
does not follow a standardized process: each \dbr\ is free to have its own means for a 
consumer to submit access requests. Since there are (respectively) $15\times$ and $25\times$ more 
deletion and opt-out requests than access requests, we hoped that submitting the most popular 
types of request would be more streamlined. However, we witnessed the same ad-hoc submission 
process for delete and opt-out requests. 
While a California resident could now use the convenient DROP platform \cite{DROP}, consumers in 
other jurisdictions (e.g. Vermont, Texas or Oregon) cannot do so and must submit individual 
requests to each \dbr.

\begin{figure}[h]
\centering
\caption{Method of Request Submission}
\vspace{0.3cm}
\fbox{\includegraphics[height=3in,width=\linewidth]{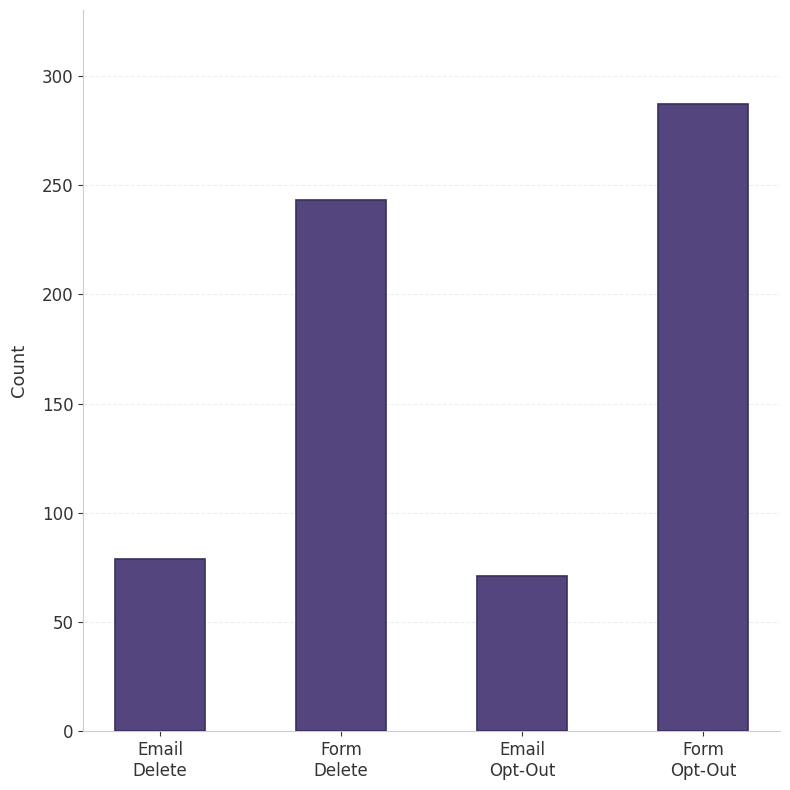}}
\label{fig:method}
\end{figure}

The majority of deletion and opt-out requests could be submitted via a form, 75\% and 80\% 
respectively. Less than a third were submitted by email. Results are shown 
in Figure \ref{fig:method}. While email request submission is faster for the consumer (since 
the same message template can be used for all email-using \dbrs), it sometimes requires a second 
step of providing more information to verify consumer identity or identify the consumer. In contrast, 
form-based submission asks for all required information at the same time: this usually means that more items
of information need to be provided to the \dbr, and each \dbr\ may ask for different types of 
personal information. In addition, forms regularly entail solving Captcha-s to prevent bots, which 
translates into an added burden for consumers \cite{bursztein2010good,searles2023empirical}.

\subsection{Identity verification process}
This section describes the details, and types, of identity verification encountered for
in submitting each request type.

\subsubsection{Deletion Requests}
\begin{figure*}[h!]
\centering
\caption{Information required for request Submission}
\vspace*{0.2cm}
\fbox{\includegraphics[width=0.8\linewidth]{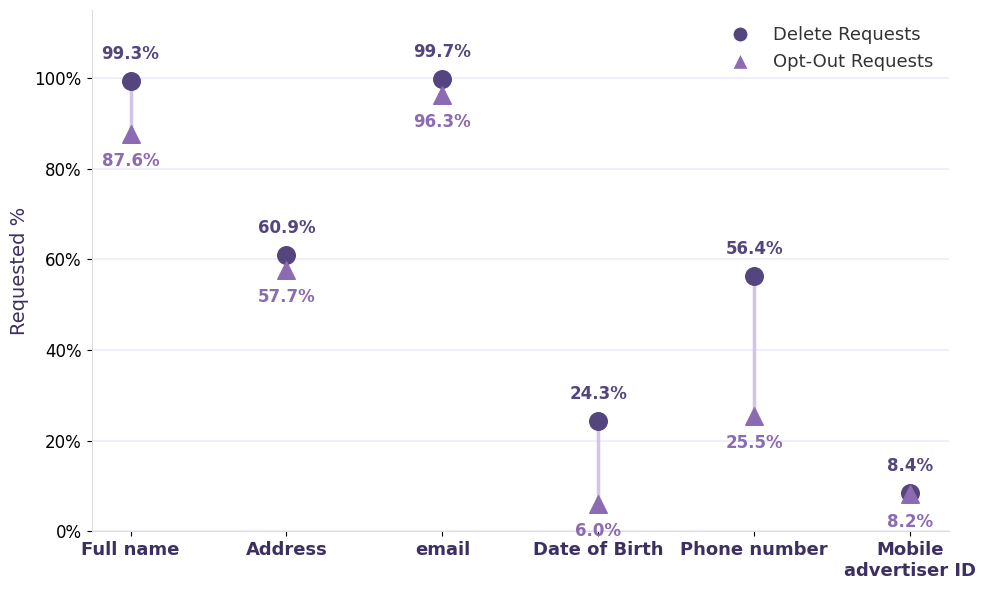}}
\label{fig:verif}
\end{figure*}

To request deletion of their personal information, consumers must submit a Verified Consumer 
Request (VCR). \dbrs\ must  
verify the identity of the consumer, and then process the VCR, i.e., delete any personal 
information relating to that consumer. To verify consumer identity, \dbrs\ ask the consumer 
to provide some items of personal information in order to match it with what is stored in their 
databases. If enough personal information matches, consumer identity is considered ``verified''
and the \dbr\  can proceed with data deletion \cite{CCR}. \dbrs\ may ask for varying types (and 
amounts of) personal information. Figure \ref{fig:verif} summarizes personal information requested by 
\dbrs\ contacted using a form. For \dbrs\ contacted via email, only the full 
name and email address of the synthetic consumer were provided.

Some \dbrs\ asked for information that we chose not to falsify because we either could not  
(e.g., device cookie values), or did not want to, do so for ethical reasons, e.g., SSNs. 

\subsubsection{Opt-Out Requests}
To opt out of sharing and further collection of their personal information, a consumer
must submit some information as part of their request, though a VCR is not required 
(see Section \ref{sec:law}): even though \dbrs\ are not supposed to request any personal information 
for verification purposes from consumers who submit opt-out requests, they can ask for 
personal information needed to identify the consumer in their databases. Despite that, We found 
that many \dbrs\ attempted to {\it verify} consumer identity during opt-out requests
by requesting additional (and sometimes sensitive) information.

\subsection{Answers and Lack Thereof}
For both deletion and opt-out requests, most \dbrs\ did not explicitly 
confirm having processed requests after a form- or email-based submission.

\subsubsection{Delete Processing\label{delpro}}
\begin{figure}[!h!]
\centering
\caption{Answers for deletion requests.}
\vspace{0.3cm}
\fbox{\includegraphics[height=5in,width=0.95\linewidth]{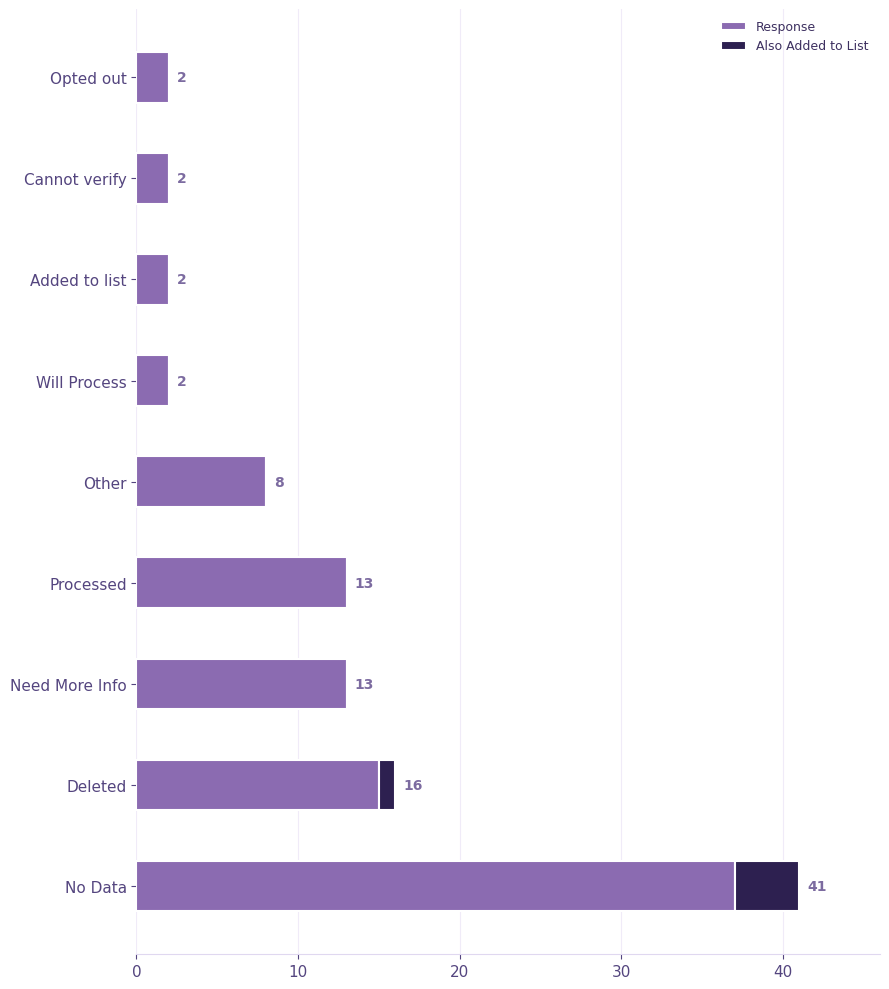}}
\label{fig:answers}
\end{figure}
As noted in Section \ref{sec:law}, \dbrs\ have 45 calendar days to inform the consumer about
the result of their deletion request. Among the 322 \dbrs\ contacted, only 97 responded with the result
of the deletion requests. We summarize their answers in Figure \ref{fig:answers}. 
Some \dbrs\ might assume that the form submission confirmation message is equivalent to the response
of confirmation for the consumer. This may be true depending on the \dbrs\ deletion process. If a \dbr\
deletes consumer's data immediately, a message stating ``your data was deleted'' suffices. Otherwise, 
the consumer is left in the dark as far as the outcome of their request. It is impossible to tell the
difference between the two cases.

\begin{figure}[!h]
\centering
\caption{Timeline for deletion requests.}
\vspace{0.3cm}
\fbox{\includegraphics[width=0.9\linewidth,height=3.5in]{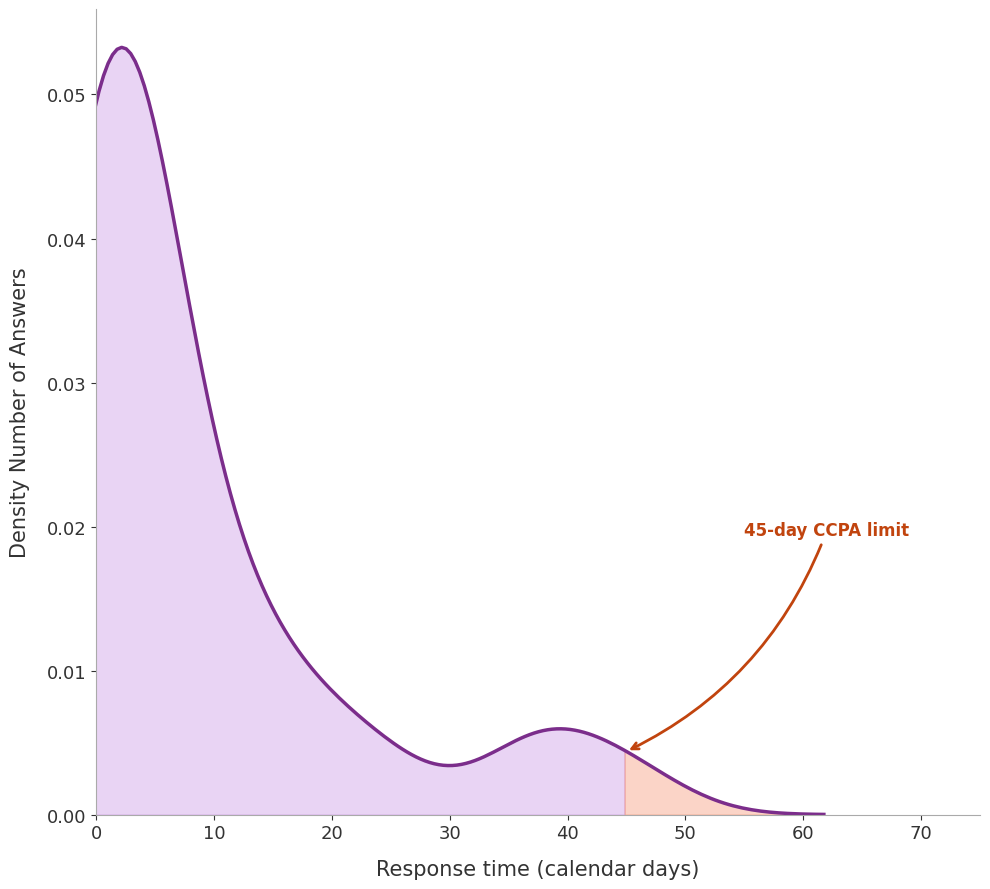}}
\label{fig:timeline}
\end{figure}

The response timeline for deletion requests was generally encouraging, with the large majority
of \dbrs\ replying within the prescribed 45-day limit (as shown in Figure
\ref{fig:timeline}), though 5\% of responding \dbrs\ exceeded it. 
Half of responding \dbrs\ replied within the first 3 calendar days.

\subsubsection{Opt-out Processing\label{optpro}}
\begin{figure}[!h]
\centering
\caption{Answers for opt-out requests.}
\vspace{0.3cm}
\fbox{\includegraphics[height=5in,width=0.95\linewidth]{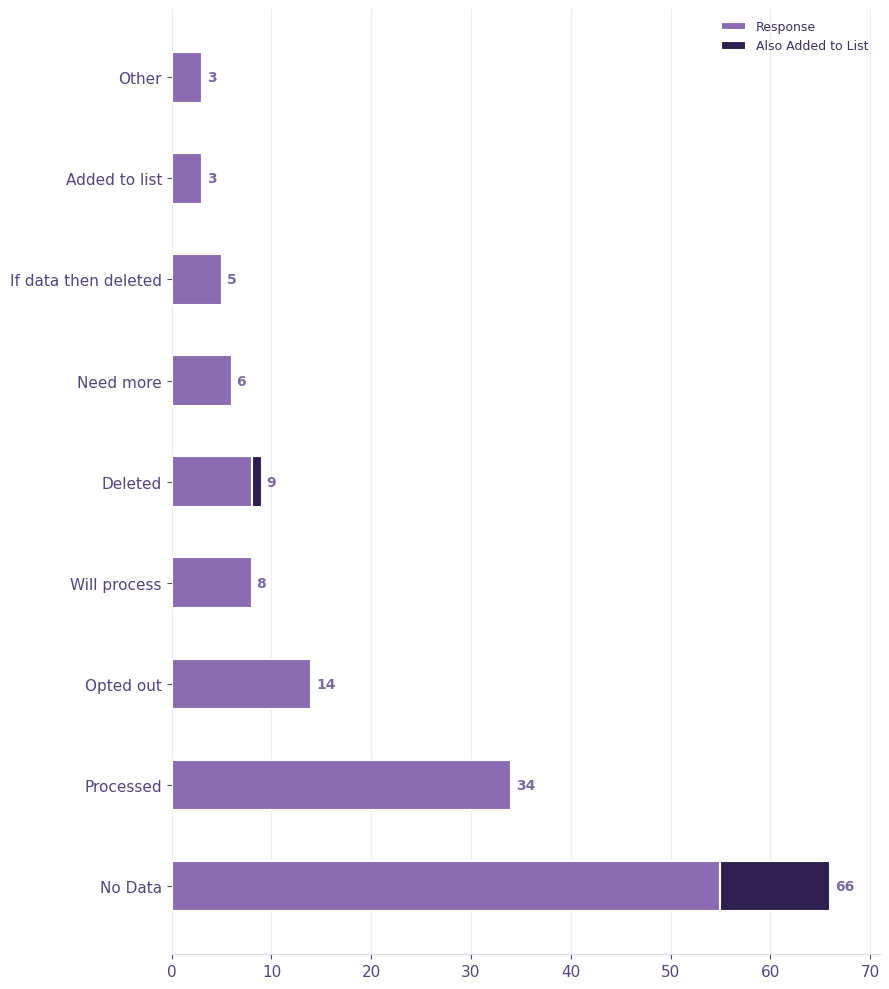}}
\label{fig:answers-optout}
\end{figure}

\dbrs\ need to act upon an opt-out request within 15 business days after its receipt. 
The law does not mandate them to directly respond to consumers. Instead, it states that a 
consumer must have a way to confirm that they have been opted out \cite{CCR}. However, without a 
response from \dbrs\ and considering the low compliance with deletion requests observed in 
this study as well as the similarly poor compliance with  access requests from \cite{van2025consumer}, 
it is unclear whether opt-out requests are actually processed appropriately. 
34 \dbrs\ confirmed, by email, that they processed the request. 66 \dbrs\ announced 
that no data was found about the consumer, of which 11 also added the consumer's 
information to an ``opt-out list''. The summary of responses appears 
in Figure \ref{fig:answers-optout}.
\begin{figure}[h]
\centering
\caption{Timeline for opt-out requests.}
\vspace{0.3cm}
\fbox{\includegraphics[width=0.9\linewidth,height=3.5in]{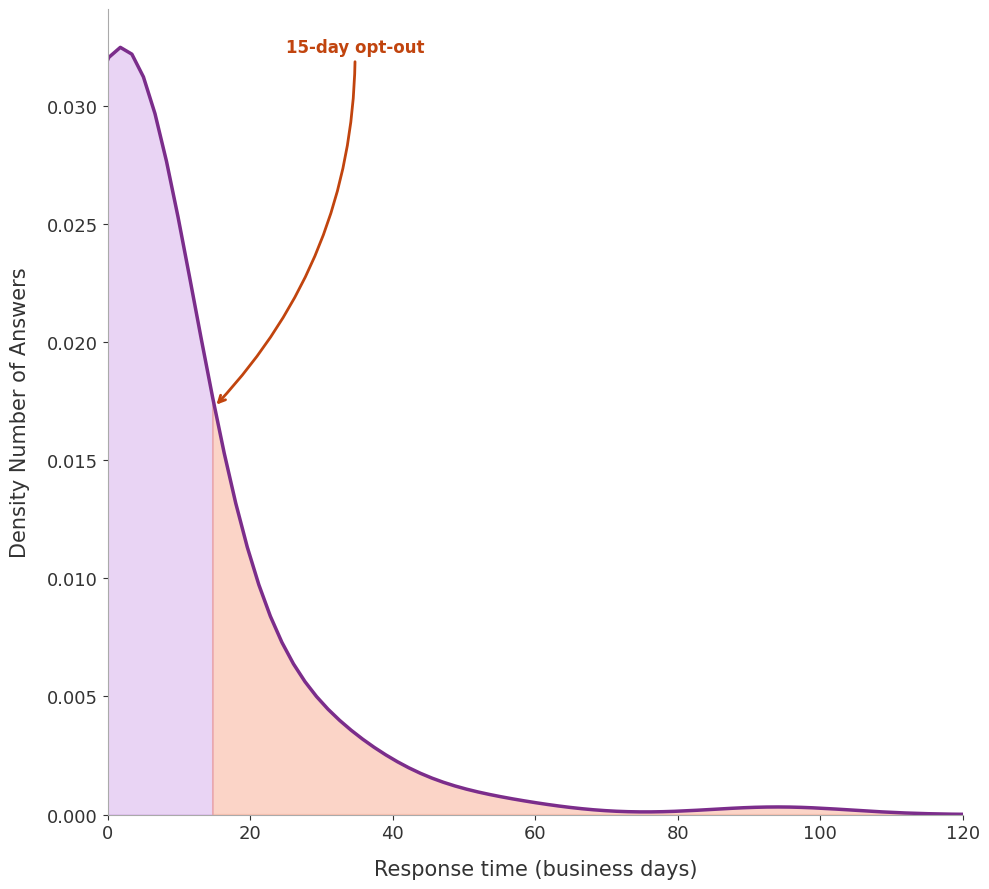}}
\label{fig:timelineOpt}
\end{figure}

The timing of responses for opt-out requests was appreciably worse than that for delete, with
a large fraction of \dbrs\ (30\%) exceeding the prescribed 15-business-day limit, 
as shown in Figure \ref{fig:timelineOpt}. Still, half of them answered within 1 business day.

\subsection{Individual Non-Compliance}
Aside from the major non-compliance trends described above, we observed the following isolated behaviors.

After submitting opt-out and deletion requests, three \dbrs\ began sending marketing emails 
to the email addresses corresponding to our synthetic identities. This practice is clearly forbidden 
by CCPA, which states that any personal information shared during a data request can only be used 
to fulfill the request, i.e. identification and verification of the consumer \cite{CCPA}. 
Any further use of the personal information is clearly forbidden.

One \dbr, responding to an opt-out request, sent a link by email which led to a simple webpage 
stating ``You have already chosen to opt-in''. Neither the email nor the webpage provided any  
information about actually opting out.

Finally, along with its confirmation of data deletion, one \dbr\  provided (via email)
a sample of its consumer records, which (alarmingly) included information about 
real people.

\subsection{Time to Submit All Requests \label{timetaken}}
Two researchers (co-authors of this study) conducted all requests submissions: 
one for each request type. Recall that a total of $358$ opt-out, and $322$ delete, 
requests were submitted. Other registered \dbrs\ could not be contacted for 
several reasons, mentioned earlier.

The submission of all opt-out requests took approximately $20$ hours, 
which corresponds to $3.3$ minutes per \dbr. In contrast, the deletion request took
about $25$ hours in total, yielding a rate of $4.6$ minutes per \dbr.
The difference of just over one minute between the two rates is mainly due to
two factors: (1) procedures for submitting deletion requests were a bit harder to 
find on the \dbrs' websites (they were often not in the same place where
deletion requests procedures were described), and (2) deletion requests required 
identity verification, while opt-out requests generally did not.

One factor that probably influenced the overall time consumed to submit requests 
is the demographic of the two researchers who performed this work. Both are Computer 
Science undergraduate students in their early- to mid-20s who are naturally 
quite tech-savvy. An average consumer is likely to take significantly longer to 
complete the same tasks.

\subsection{Comparison With a Prior Study of Data Access Requests}
As mentioned earlier, 
a closely related study was recently conducted by Van Kempen et al. \cite{van2025consumer}. 
It focused on data access requests submitted under CCPA to all California-registered \dbrs\ in late
2024  At that time, $543$ \dbrs\ were registered and requests were made to $454$ of those. 
In contrast, we submitted opt-out and deletion requests to $358$ and $322$ \dbrs, respectively. 
This represents substantial differences in terms of numbers of contacted \dbrs.

Since \cite{van2025consumer} used a real person's (one of its authors') information to 
submit data access requests, more \dbrs\ could be contacted. However, our study
encountered more limitations since we used synthetic identities and thus could not (or did not
want to) fabricate certain information, e.g. cookie values or SSN-s.
We believe that this accounts for most of the gap in request submission numbers.

Compared to \cite{van2025consumer} where $48.9$\% of \dbrs\ provided form submissions, 
many more \dbrs\ did so in this study: $75$\% and $80$\%. We believe that there were
two reasons for this:
\begin{enumerate}
    \item Requests to delete/opt-out are easier to automate than requests to access. 
    Therefore, use of a form streamlines the process for \dbrs.
    \item Over the past couple of years, more \dbrs\ adopted form 
    submissions for all types of consumer requests.
\end{enumerate}
Beyond the numbers of contacted \dbrs, response rates in our study were also
lower than that in the prior study of access requests. This is likely because 
\dbrs\ assume that a form submission confirmation page/message for a delete/opt-out 
request represents a response, and thus do not separately confirm receipt
of a request.

In terms of responses, \dbrs\ followed the same answer timeline pattern
in both studies: a vast majority responded 
within the first few days, and very few \dbrs\ answered late.

\subsection{Answers to Research Questions}
Recall the research questions posed in Section \ref{intro}. We now 
describe the extent to which the study answered them. \\

\noindent\textbf{RQ1: What are the differences, if any, 
in the submission processes of opt-out and delete requests?}  
The former involve less consumer burden and are thus faster to submit 
since they require no identity verification. In general, opt-out requests needed 
less personal information (see Figure \ref{fig:verif}). However, a significant percentage of
\dbrs\ still asked to verify consumer identity, thus not complying 
with CCPA. Also, deletion requests are more final, since \dbrs\ can still 
keep (previously collected) information about a given consumer after an opt-out request. 
Within each type of request, we found that most \dbrs rely on submission forms for 
opt-out or deletion requests rather than email. These forms require providing a 
different set of PII to each \dbr.\\
    
\noindent\textbf{RQ2: To what extent do brokers comply with CCPA in terms of
of delete and opt-out requests?}
As described is Section \ref{sec:compliance}, significant non-compliance was observed. 
The majority of \dbrs\ did not respond to our deletion requests, and many did 
not confirm our opt-out requests. Furthermore, potential acts of non-compliance 
were common. Results are summarized in Table \ref{tab:res-sum}.\\

\noindent\textbf{RQ3: What is the overall consumer burden involved in submitting 
each request type to  all \dbrs?} 
Since every \dbr\ has its own delete/opt-out request submission process 
(see Section \ref{sec:nostandard}), a consumer needs to access every \dbr's webpage and 
find request submission directions. With well over $300$ \dbrs, this represents a very heavy 
burden (in terms of time and effort) for an average consumer. It is also highly repetitive
and thus extremely tedious. While California consumers can now
use the DROP \cite{DROP} portal, those in other jurisdictions can either contact each \dbr\
separately or pay a third-party service to so on their behalf.

\section{Study Limitations \label{limits}}
Since we used synthetic identities, we cannot be sure whether \dbrs\ actually complied
with our requests. \dbrs\ could have ignored requests if they deemed them fraudulent or
otherwise inauthentic. In the best case, \dbrs\ should have recorded  our opt-out requests even
though they should not have found any information in their databases corresponding to the
fictitious individuals. In other words, they should have kept ``state'' to prevent themselves 
from collecting any information about that individual in the future. 

As far as deletion requests, the best outcome would have been an 
explicit notification by each \dbr\ about their inability to find any information 
for the fictitious individual. However, this is a fundamental limitation of our study: 
although it uncovered various types of non-compliance, it did not evaluate the actual 
{\it efficacy} of submitting delete and opt-out requests. Doing that would 
require using real identities, which would have been problematic.

\section{Future Work \label{future}}
The study presented here is by no means the ``end of the road.'' 
The new California DROP system mandates \dbrs\ to explicitly communicate 
the status of each request to the requesting consumer. This feature, if 
treated in good faith (by \dbrs) and implemented
well by CPPA, would be highly beneficial to consumers. 
One intuitive direction for future work is to wait about 6-12 months after August 2026 (when \dbrs\
were first mandated to actually participate in the CPPA's DROP portal) and assess its efficacy.

Another direction is to repeat the study presented in this paper with identities that are
neither synthetic nor those of the researchers conducting the study. The main idea would be to 
evaluate deletion and opt-out (by submitting requests as we did) and then later 
verify that those requests were acted upon by submitting data access requests. We obviously
could not do the latter, since we used non-existent identities. However, using identities of real
people can be problematic, e.g., recruitment and remuneration, as well as potential ethical
issues. An alternative is to use identities of dead people, though this is likely even more problematic. 

As mentioned in Section \ref{timetaken}, the two researchers who submitted both request types 
are not representative of the general population, due to being quite young and technologically 
adept. A more realistic assessment of overall time and effort would use a representative 
population sample. 

Finally, in the long-term, a more frictionless user interface to \dbrs\ should be developed.
While the CPPA's DROP portal is a step in the right direction, it is not privacy agile.
A consumer must enter lots of personal information into the portal. Also, DROP operates in 
an asynchronous manner, since \dbrs\ are only required to fetch accumulated (on the portal) 
delete and opt-out requests with a frequency of at most 45 days. It is not hard to imagine
a system that offers more privacy for consumers and operates in a real-time fashion.

\section{Conclusions \label{conc}}
While current regulations, exemplified by CCPA, appear to give consumers more control
over their data, exercising one’s rights is still a  highly laborious and opaque 
process. As the results of the study described in this paper show, one must submit 
many types of forms (and/or emails) and what comes back (if anything at all)
is mostly in the form of generic text. Deletion is harder than
opt-out and sometimes requires sending a considerable amount of
additional personal information to \dbrs.

\raggedbottom

\pagebreak
\eject
\newpage

\bibliographystyle{plainurl}
\bibliography{references}

\end{document}